\documentclass[conference]{IEEEtran}
\IEEEoverridecommandlockouts
\usepackage{booktabs}
\usepackage{cite}
\usepackage{amsmath,amssymb,amsfonts}
\usepackage{algorithmic}
\usepackage{algorithm}
\usepackage{amsmath}
\usepackage{graphicx}
\usepackage{textcomp}
\usepackage{xcolor}
\usepackage{subfigure}
\def\BibTeX{{\rm B\kern-.05em{\sc i\kern-.025em b}\kern-.08em
    T\kern-.1667em\lower.7ex\hbox{E}\kern-.125emX}}
\begin{document}

\title{Learning to Jointly Optimize Antenna Positioning and Beamforming for Movable Antenna-Aided Systems\\

}

\author{
    \IEEEauthorblockN{Yikun Wang\IEEEauthorrefmark{1}\IEEEauthorrefmark{2}, Yang Li\IEEEauthorrefmark{2}, Zeyi Ren\IEEEauthorrefmark{1}, Jingreng Lei\IEEEauthorrefmark{1}\IEEEauthorrefmark{2}, Yik-Chung Wu\IEEEauthorrefmark{1}, and Rui Zhang\IEEEauthorrefmark{3}}
    
    \IEEEauthorblockA{\IEEEauthorrefmark{1}Department of Electrical and Electronic Engineering, The University of Hong Kong, Hong Kong}

    \IEEEauthorblockA{\IEEEauthorrefmark{2}School of Computing and Information Technology,
Great Bay University, Dongguan, China}

\IEEEauthorblockA{\IEEEauthorrefmark{3}Department of Electrical and Computer Engineering, National University of Singapore, Singapore}


    \IEEEauthorblockA{Emails: \{ykwang, renzeyi, leijr, ycwu\}@eee.hku.hk, liyang@gbu.edu.cn, elezhang@nus.edu.sg
}
}

\maketitle

\maketitle

\begin{abstract}
The recently emerged movable antenna (MA) and fluid antenna technologies offer promising solutions to enhance the spatial degrees of freedom in wireless systems by dynamically adjusting the positions of transmit or receive antennas within given regions. In this paper, we aim to address the joint optimization problem of antenna positioning and beamforming in MA-aided multi-user downlink transmission systems. This problem involves mixed discrete antenna position and continuous beamforming weight variables, along with coupled distance constraints on antenna positions, which pose significant challenges for optimization algorithm design. To overcome these challenges, we propose an end-to-end deep learning framework, consisting of a positioning model that handles the discrete variables and the coupled constraints, and a beamforming model that handles the continuous variables. Simulation results demonstrate that the proposed framework achieves superior sum rate performance, yet with much reduced computation time compared to existing methods.
\end{abstract}


\section{Introduction}
Movable antenna (MA) \cite{10318061} and fluid antenna (FA) \cite{9264694} technologies have garnered significant attention in recent years. Unlike traditional fixed-position antennas, MAs and FAs can dynamically change their positions within a designated area. This flexibility allows antennas to move away from deep-fading positions, thereby improving the channel state and creating more favorable conditions for wireless transmissions. 

\let\thefootnote\relax\footnotetext{The work of Yikun Wang, Yang Li, and Jingreng Lei was supported in part by the National Natural Science Foundation of China (NSFC) under Grant 62571086, in part by Guangdong Basic and Applied Basic Research Foundation under Grant 2025A1515011658, and in part by Guangdong Research Team for Communication and Sensing Integrated with Intelligent Computing (Project No. 2024KCXTD047).}

\let\thefootnote\relax\footnotetext{The codes for generating the simulation results in this paper are available at: https://github.com/YikunWang-EEE/RL-MA-joint-APS-and-BF.}

The MA-aided system achieves its optimum performance through antenna positioning\cite{10243545}. Prior research has explored the antenna positioning problem in continuous space \cite{10354003,10243545,ycjin,C_FP_MA}. They aim to maximize various utility functions, e.g., sum rate, minimum signal-to-interference-plus-noise ratio (SINR), by optimizing the MA positions within a continuous designated area. Additionally, some studies have focused on the discrete antenna positioning problem\cite{graph,Graph-based-Security,derrick,6DMA}, where the designated area is sampled into multiple position points. As such, the original positioning optimization in continuous space is transformed to a sampling point (SP) selection problem. This approach simplifies the hardware implementation but brings additional challenges for optimization algorithm design: the problem becomes combinatorial, and constrained by the coupled distance requirements between antennas. In \cite{graph,Graph-based-Security}, graph theory-based approaches were proposed by modeling the problem as a fixed-hop shortest path problem. However, they are only applicable for the single-user scenario, limiting their applicability in multi-user systems when inter-user interference exists. While \cite{derrick} proposed a generalized Bender’s decomposition method to find the optimal antenna positions and beamformers that minimize the transmit power while ensuring the minimum SINR, this method suffers from exponential complexity in the worst case, which is practically hard for real-time implementation. Moreover, discrete antenna position and rotation joint design has been studied in six-dimensional movable antenna (6DMA) aided systems\cite{6DMA}. 

Deep learning (DL) has been successfully applied to solve numerous optimization problems in communication systems, owing to its capacity to model complex functions and its affordable real-time deployment compared to iterative-optimization algorithms. However, the joint design of discrete positioning and beamforming poses distinctive challenges. First, the underlying optimization problem encompasses mixed discrete and continuous variables, where the discrete variables induce zero-gradient issues during backpropagation, rendering them difficult to handle within neural network (NN) architectures. Furthermore, the NN is required to generate discrete positioning solutions that satisfy the coupled distance constraints, yet ensuring consistent fulfillment of these constraints in the NN's outputs remains inherently challenging.

To fill this gap, we propose a novel DL framework comprising two distinct NN models that handle positioning and beamforming sequentially. The positioning NN incorporates a graph neural network (GNN)-and-attention-based encoder-decoder architecture, which outputs the joint distribution of the positioning solutions. Through a judicious mask design, the generated positioning solutions are guaranteed to satisfy the coupled distance constraints, without any post-processing. Then, a GNN-based beamforming NN is employed to optimize the beamformers, leveraging an optimal solution structure to simplify the mapping to be learned. Furthermore, we introduce an end-to-end training algorithm, where the positioning NN and the beamforming NN are jointly trained in an unsupervised manner. Simulation results demonstrate that the proposed DL framework achieves superior sum rate performance than all baselines, while exhibiting significantly faster inference speed.

\section{System Model}
Consider an MA-aided downlink system with one base station (BS) serving $K$ user equipments (UEs). The BS is equipped with $M$ MAs and each UE is equipped with a single fixed-position antenna. The positions of the $M$ MAs can be adjusted simultaneously within a predefined two-dimensional rectangular area, which is divided into $N\gg M$ SPs for placing the $M$ MAs. The coordinate of the $n$-th SP is denoted by $\mathbf{p}_n = [x_n, y_n]^T$, where $x_n$ and $y_n$ represent the coordinates along the x-axis and y-axis, respectively, in the Cartesian coordinate system.
Let $\tilde{\mathbf{h}}_k \triangleq [\tilde{h}_{1k},\tilde{h}_{2k},\dots,\tilde{h}_{Nk}]^T$ denote the channel between the $N$ SPs and UE $k \in \mathcal{K}\triangleq\{1,2,\dots,K\}$, which is assumed to be known\cite{graph}. 

Define the selected SP set as $\mathcal{A}\triangleq\{a_1,a_2,\dots,a_{M}\}$, where $a_m\in\mathcal{N}\triangleq\{1,2,\dots,N\}$. The received SINR of the $k$-th UE is given by

\begin{equation} 
\text{SINR}_{k} =\frac{\left | \mathbf{h}_{k}(\mathcal{A})^{H}\mathbf{w}_{k} \right |^{2}}{\sum_{l=1,l\ne k}^{K}\left | \mathbf{h}_{k}(\mathcal{A})^{H}\mathbf{w}_{l} \right |^{2}+\sigma^2_k},
\label{SINR}
\end{equation}
where $\mathbf{h}_k(\mathcal{A})\in \mathbb{C}^{M}$ is the channel between the selected $M$ SPs and the $k$-th UE, $\mathbf{w}_{k} \in \mathbb{C}^{M}$ represents the transmit beamformer for the $k$-th UE, and $\sigma^2_k$ is the power of the additive-white-Gaussian-noise. In this paper, we aim to maximize the sum rate by jointly optimizing $\mathcal{A}$ and $\mathbf{w}\triangleq[\mathbf{w}_1^T,\mathbf{w}_2^T,\dots,\mathbf{w}_{K}^T]^T$, which is formally formulated as
\begin{subequations}
\label{optim}
\begin{align}
& \max _{\mathcal{A},\mathbf{w}}&&R(\mathcal{A},\mathbf{w};\mathbf{h})\triangleq\sum_{k=1}^{K}\log_2\left( 1 +\text{SINR}_k\right)&
\label{obj_of_SR} \\
& ~~\text{s.t.}&&a_m\in \mathcal{N}, ~~\forall m \in \mathcal{M}\triangleq\{1,2,\dots,M\},& \label{discrete_constraint}\\ 
&& &\|\mathbf{p}_{a_m}-\mathbf{p}_{a_{m'}}\|\ge d_{\text{min}},~\forall m,m'\in\mathcal{M}, ~m\ne m',&\label{distance_constraint}\\
&&&\sum_{k=1}^{K}\left\|\mathbf{w}_{k}\right\|^{2} \leq P_{\text{max}},\label{power}&
\end{align}
\end{subequations}
where $\mathbf{h} \triangleq [\tilde{\mathbf{h}}_1^T,\tilde{\mathbf{h}}_2^T,\dots,\tilde{\mathbf{h}}_{K}^T,\mathbf{p}_1^T,\mathbf{p}_2^T,\dots,\mathbf{p}_{N}^T]^T$ denotes the known system parameters, $d_{\min}$ is the minimum distance between two MAs, and $P_{\max}$ is the maximum transmit power of the BS.

Problem (\ref{optim}) is combinatorial, making it NP-hard to find the optimal solution. Furthermore, developing a DL-based approach poses two key challenges: \textit{(a)} the coupling between discrete and continuous variables; and \textit{(b)} the highly non-convex discrete constraints in (\ref{distance_constraint}). In the subsequent section, we introduce a novel end-to-end DL framework that addresses these issues by handling the variables $\mathcal{A}$ and $\mathbf{w}$ sequentially through two distinct NN models.

\section{Proposed DL Framework}
The proposed end-to-end DL framework comprises two NN models: a positioning NN and a beamforming NN. Specifically, the positioning NN takes the system parameter $\mathbf{h}$ as the input, and outputs the selected SP set $\mathcal{A}$, i.e., $\mathcal{A}=\mathcal{F}_{\text{p}}(\mathbf{h})$. Then, the beamforming NN takes $\mathcal{A}$ and $\mathbf{h}$ as the input, and outputs the beamformer $\mathbf{w}$, i.e., $\mathbf{w}=\mathcal{F}_{\text{w}}(\mathcal{A},\mathbf{h})$. These models are cascaded and trained jointly to provide near-optimal solutions while satisfying the constraints at the same time. We then detail the design of $\mathcal{F}_{\text{p}}(\cdot)$, $\mathcal{F}_{\text{w}}(\cdot,\cdot)$, and the joint training algorithm, respectively.

\subsection{Design of $\mathcal{F}_{\text{p}}(\cdot)$}
Since $\mathcal{A}$ consists of discrete variables, to avoid the zero-gradient due to a hard decision, we treat the elements in $\mathcal{A}$ as random variables, and strive to learn the conditional probability of $\mathcal{A}$ given any $\mathbf{h}$. Furthermore, due to the coupled distance constraints in (\ref{distance_constraint}), each element of $\mathcal{A}$ depends on each other. Consequently, we factorize this conditional probability as
\begin{equation}
p(\mathcal{A} \mid {\mathbf{h}})= \prod_{t=1}^{M} p(a_{m(t)} \mid \mathcal{A}_{t-1}, \mathbf{h}),
\label{factorization}
\end{equation}
where $a_{m(t)}$ denotes the selected SP at the $t$-th step, $\mathcal{A}_{t-1}\triangleq\{a_{m(t')}\}_{t'=1}^{t-1}$ represents all the SPs that have already been selected up to the ($t-1$)-th step, and $\mathcal{A}_{0}\triangleq\emptyset$.  Next, we construct an encoder-decoder model to learn the conditional probability (\ref{factorization}), which also guarantees the coupled distance constraints in (\ref{distance_constraint}).

\textit{1) Design of the encoder: }The encoder $\mathcal{G}_{\text{E}}(\cdot)$ aims to build the mapping function from the system parameter $\mathbf{h}$ to the embeddings of all SPs denoted as $\mathbf{R}\triangleq[\mathbf{r}_1,\mathbf{r}_2,\dots,\mathbf{r}_{N}]\in \mathbb{R}^{n_{\text{emb}} \times N}$, capturing their suitability of antenna positioning, where $n_{\text{emb}}$ denotes the embedding dimension. To this end, we model the MA-aided system as a graph, which includes two types of nodes, i.e., $K$ UE-nodes
and $N$ SP-nodes, and there exists an edge between each UE-node and each SP-node. We first show a desired permutation invariance (PI)-permutation equivariance (PE) property of $\mathcal{G}_\text{E}(\cdot)$ in the following property.

\textit{Property 1 (PI-PE Property of $\mathcal{G}_{\text{E}}(\cdot)$):} Let $\hat{h}_{nk} \triangleq \tilde{h}_{\pi_2(n)\pi_1(k)} $, $\hat{\bf{p}}_n=\mathbf{p}_{\pi_2(n)}$, where $\pi_1(\cdot)$ and $\pi_2(\cdot)$ are permutations of the indices
in $\mathcal{K}$ and $\mathcal{N}$, respectively. The mapping function $\mathcal{G}_{\text{E}}(\cdot)$ satisfies
\begin{equation}
\begin{aligned}
&{\hat{\mathbf{R}}}=\mathcal{G}_{\text{E}}(\Re\{\hat{\mathbf{h}}\},\Im\{\hat{\mathbf{h}}\}), \\
&\forall\pi_{1}(\cdot):\mathcal{K}\to\mathcal{K},~~\forall\pi_{2}(\cdot):\mathcal{N}\to\mathcal{N},
\end{aligned}
\end{equation}
if and only if $\mathbf{R}=\mathcal{G}_{\text{E}}(\Re\{\mathbf{h}\},\Im\{\mathbf{h}\})$ holds, where $\hat{\mathbf{h}}\triangleq[\hat{\mathbf{h}}_{\pi_1(1)}^T, \hat{\mathbf{h}}_{\pi_1(2)}^T,\dots,\hat{\mathbf{h}}_{\pi_1(K)}^T, \mathbf{p}_{\pi_2(1)}^T, \mathbf{p}_{\pi_2(2)}^T,\dots,\mathbf{p}_{\pi_2(N)}^T]^T$, $\hat{\mathbf{h}}_k \triangleq [h_{\pi_2(1)k}, h_{\pi_2(2)k},\dots, h_{\pi_2(N)k}]^T$, and $\hat{\mathbf{R}}\triangleq[\mathbf{r}_{\pi_2(1)},\mathbf{r}_{\pi_2(2)},\dots,\mathbf{r}_{\pi_2(N)}]$.

To ensure \emph{Property 1}, we design an edge-node GNN (ENGNN) \cite{engnn} as the foundation architecture of $\mathcal{G}_{\text{E}}(\cdot)$, where the $k$-th UE-node feature and $n$-th SP-node feature of the $l_{\text{E}}$-th hidden layer are expressed as $\mathbf{f}^{[l_{\text{E}}]}_{\text{UE},k}$ and $\mathbf{f}^{[l_{\text{E}}]}_{\text{SP},n}$, and the corresponding edge feature is represented by $\mathbf{e}^{[l_{\text{E}}]}_{nk}$.
The edge features and SP-node features are initialized as
\begin{subequations}
\begin{align}
&\mathbf{e}_{nk}^{[0]}=\text{MLP}^{\text{E}}_{1}\left(\Re\{\tilde{h}_{nk}\},\Im\{\tilde{h}_{nk}\}\right),~ \forall k \in \mathcal{K},n \in \mathcal{N},\label{MAEnPre} \\
&\mathbf{f}^{[0]}_{\text{SP},n}=\text{MLP}^{\text{E}}_{2}\left(\mathbf{p}_n\right),~~\forall n\in \mathcal{N},
\end{align}
\end{subequations}
where $\text{MLP}_1^{\text{E}}(\cdot)$ and $\text{MLP}_2^{\text{E}}(\cdot)$ are 2 multi-layer perceptrons (MLPs), and the UE-node features $\{\mathbf{f}^{[0]}_{\text{UE},k}\}_{k\in\mathcal{K}}$ are initialized as zero.
\addtolength{\topmargin}{0.040in}
After that, the encoder updates its node and edge features by a node-edge update mechanism, which is expressed as
\begin{equation}
\label{ENGNNUE}
\begin{aligned}
\mathbf{f}_{\text{UE},k}^{[l_{\text{E}}]}= & \text{MLP}_{2}^{[l_{\text{E}}]}  \left(\mathbf{f}_{\text{UE},k}^{[l_{\text{E}}-1]}, \frac{1}{N}\sum_{n=1}^{N}\text{MLP}_{1}^{[l_{\text{E}}]}\left(\mathbf{f}_{\text{SP},n}^{[l_{\text{E}}-1]}, \mathbf{e}_{nk}^{[l_{\text{E}}-1]}\right)\right), \\
&\forall k \in \mathcal{K},
\end{aligned}
\end{equation}
\begin{equation}
\label{ENGNNSP}
\begin{aligned}
\mathbf{f}_{\text{SP},n}^{[l_{\text{E}}]}= & \text{MLP}_{4}^{[l_{\text{E}}]} \left(\mathbf{f}_{\text{SP},n}^{[l_{\text{E}}-1]}, \frac{1}{K}\sum_{k=1}^{K}\text{MLP}_{3}^{[l_{\text{E}}]}\left(\mathbf{f}_{\text{UE},k}^{[l_{\text{E}}-1]}, \mathbf{e}_{nk}^{[l_{\text{E}}-1]}\right)\right), \\
& \forall n \in \mathcal{N},
\end{aligned}
\end{equation}
\begin{equation}
\label{ENGNNEdge}
\begin{aligned}
&\mathbf{e}_{nk}^{[l_{\text{E}}]}=\text{MLP}_{7}^{[l_{\text{E}}]} \left(\mathbf{e}_{nk}^{[l_{\text{E}}-1]}, \frac{1}{N}\sum_{n'=1}^{N}\text{MLP}_{5}^{[l_{\text{E}}]}\left(\mathbf{e}^{[l_{\text{E}}-1]}_{n'k},\mathbf{f}_{\text{UE},k}^{[l_{\text{E}}-1]}\right),\right.\\
&~~~~~\left.\frac{1}{K}\sum_{k'=1}^{K}\text{MLP}_{6}^{[l_{\text{E}}]}\left(\mathbf{e}_{nk'}^{[l_{\text{E}}-1]},\mathbf{f}_{\text{SP},n}^{[l_{\text{E}}-1]}\right)\right), \forall k \in \mathcal{K},n \in \mathcal{N},
\end{aligned}
\end{equation}
where $\text{MLP}_{1}^{[l_{\text{E}}]}(\cdot)$ - $\text{MLP}_{7}^{[l_{\text{E}}]}(\cdot)$ are 7 MLPs. After $L_{\text{E}}$ layers' update, the final SP-node feature $\mathbf{f}^{[L_{\text{E}}]}_{\text{SP},n}$ serves as the required embedding $\mathbf{r}_{n}$.

\begin{figure}[htbp]
\centerline{\includegraphics[scale=0.148]{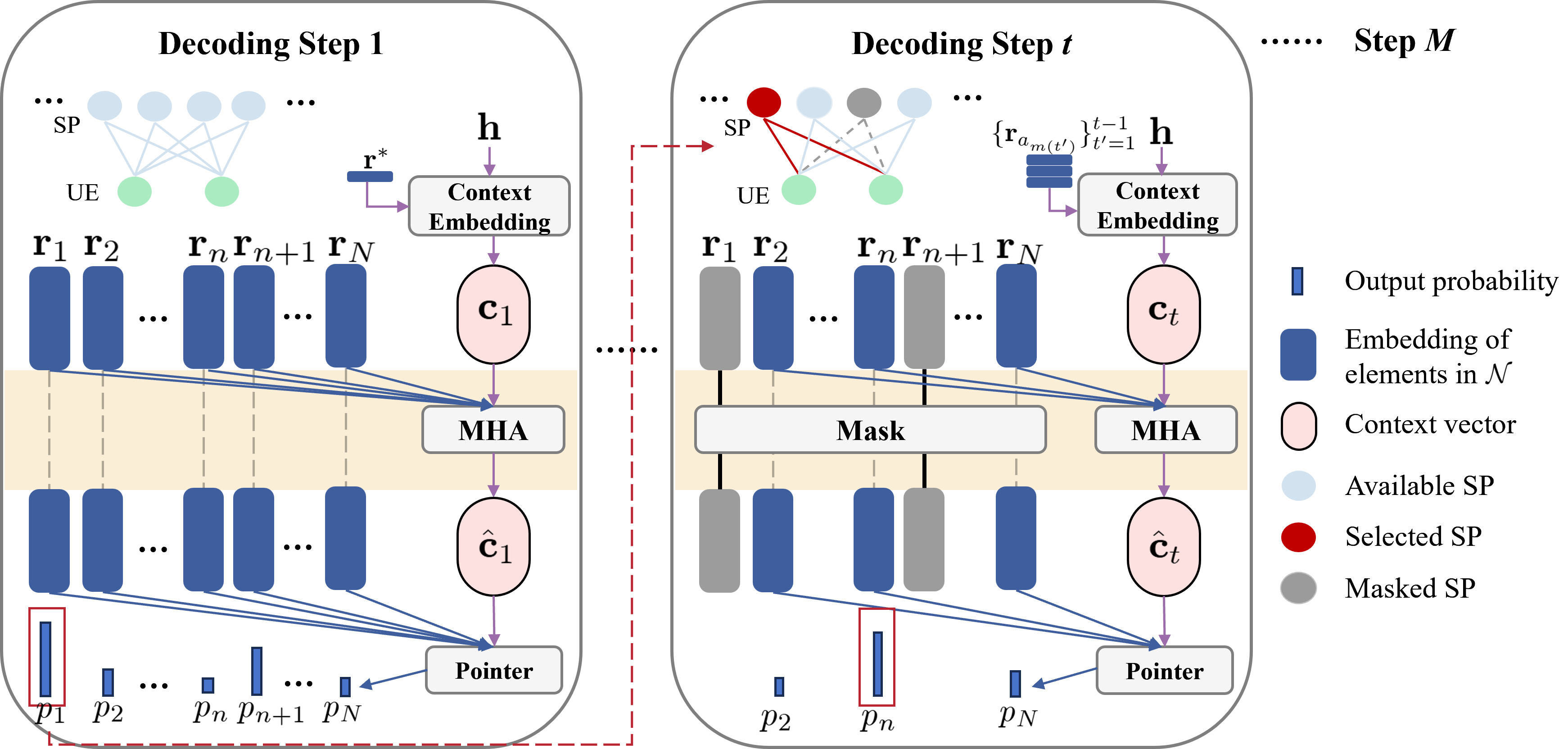}}
\caption{The illustration of the decoding process.}
\label{fig2}
\end{figure}

\textit{2) Design of the decoder: }
The decoder obtains the conditional probability $p(\mathcal{A} \mid \mathbf{h})$ in $M$ steps. At the $t$-th step, the decoder firstly extracts the current system state information, i.e., $\mathcal{A}_{t-1}$ and $\mathbf{h}$, by a \emph{context embedding}. Then, the \emph{multi-head attention (MHA)} is applied between the context vector and $\mathbf{R}$, to further extract deeper-level features. Finally, a \emph{Pointer} \cite{bello*2017neural} module is applied to capture the compatibilities between the context vector and $\mathbf{R}$, so as to output the conditional probability $p(a_{m(t)}|\mathcal{A}_{t-1},\mathbf{h} )$. The next SP $a_{m(t)}$ is sampled from this conditional probability. The aforementioned step is repeated for $M$ times, until $\mathcal{A}$ is fully determined. The decoding process is illustrated in Fig. \ref{fig2}, where the detailed steps are presented as follows.

\textbf{Context Embedding:} When $t\ge2$, the context embedding is computed as
\begin{equation}
\begin{aligned}
\mathbf{c}_{t}&=
\text{MLP}_1^{\text{X}}\left(\frac{1}{t-1}\sum\limits_{t'=1}^{t-1}\text{MLP}_4^{\text{X}}(\mathbf{r}_{a_{m(t')}}),\right. \\
& \left. \frac{1}{N}\sum_{n=1}^{N}\text{MLP}_2^{\text{X}}\left(\mathbf{p}_{n},\frac{1}{K}\sum_{k=1}^{K}\text{MLP}^{\text{X}}_{3}\left(\Re\{{\tilde{h}_{nk}}\},\Im\{{\tilde{h}_{nk}}\}\right)\right)\right),
\end{aligned}
\end{equation}
where $\mathbf{c}_t$ is a $d_{\text{h}}$-dimensional vector, and $\text{MLP}^{\text{X}}_1(\cdot,\cdot)$-$\text{MLP}^{\text{X}}_4(\cdot)$ are 4 MLPs. When $t=1$, as $\mathcal{A}_{t-1}=\emptyset$, we use a trainable parameter $\mathbf{r}^{*} \in \mathbb{R}^{d_{\text{h}}}$ to replace the first input term of  $\text{MLP}^{\text{X}}_1(\cdot,\cdot)$.

\textbf{MHA:} For ease of expression, the index $t$ is omitted. Let $N_{\text{h}}$ denote the number of attention heads. The query $ \mathbf{q}^{[h]}$, the key $ \mathbf{k}_{n}^{[h]}$, and the value $ \mathbf{v}_{n}^{[h]}$ at the $h$-th head are computed as follows:
\begin{subequations}
\begin{align}
& \mathbf{q}^{[h]}=\mathbf{W}_{\text{Q}}^{[h]} \mathbf{c},  \quad \forall h=1,2,\dots,N_\text{h},\\
&\mathbf{k}_{n}^{[h]}=\mathbf{W}_{\text{K}}^{[h]} \mathbf{r}_{n},  \quad\forall n\in\mathcal{N} ,~~\forall h=1,2,\dots,N_\text{h}, \\
& \mathbf{v}_{n}^{[h]}=\mathbf{W}_{\text{V}}^{[h]} \mathbf{r}_{n},\quad\forall n\in\mathcal{N} ,~~\forall h=1,2,\dots,N_\text{h},
\end{align}
\end{subequations}
where $\mathbf{q}^{[h]}$, $ \mathbf{k}_{n}^{[ h]}$, and $ \mathbf{v}_{n}^{[ h]}$ are all $d_\text{v}$-dimensional vectors, $d_\text{v} = {d_\text{h}}/{N_\text{h}}$, and $\mathbf{W}_{\text{Q}}^{[h]} $, $\mathbf{W}_{\text{K}}^{[h]} $, and $\mathbf{W}_{\text{V}}^{ [h]}$ are trainable matrices. 

After that, we compute the relevance scores between the query and all keys. For each decoding step $t$, we define an available SP set as $\bar{\mathcal{N}}\triangleq\left\{n\in\mathcal{N}\left| \left\|\mathbf{p}_{n}-\mathbf{p}_{a_{m(t')}}\right\|\ge d_{\text{min}},\forall t'<t\right\}\right.$. As such, the relevance score with respect to the $n$-th SP is obtained as
\begin{equation}
\begin{split}
u_{n}^{[ n_\text{h}]} &= \begin{cases}
\frac{\left(\mathbf{q}^{[ n_\text{h}]} \right)^{T}\mathbf{k}_{n}^{[ n_\text{h}]}}{\sqrt{d_\text{v}}} & ,\text{ if } n\in \bar{\mathcal{N}}, \\
-\infty & ,\text{ otherwise},
\end{cases} \\
\end{split}
\end{equation}
where we mask the SPs that have already been chosen or violate the distance constraint (\ref{distance_constraint}) by setting their score as minus infinity. Consequently, we update the context vector by aggregating the value of each SP with the normalized relevance score serving as the corresponding weight:
\begin{equation}
\hat{\mathbf{c}} = \sum_{h=1}^{N_\text{h}}\left[\mathbf{W}_\text{O}^{[ h]}\left(\sum_{n=1}^{N}\frac{e^{u_{n}^{[h]}}}{\sum_{n'=1}^{N}e^{u_{n'}^{[ h]}}}\mathbf{v}_{n}^{[h]}\right)\right],
\end{equation}
where the trainable matrix $\mathbf{W}_\text{O}^{[h]} \in \mathbb{R}^{d_\text{h}\times d_\text{v}}$ maps the multiple aspects generated by MHA back to a unified space.

\textbf{Pointer:}
In the \emph{Pointer} module, we first adopt a single-head attention mechanism to compute the compatibilities between the updated context vector $\hat{\mathbf{c}}$ and all the SPs as
\begin{equation}
u_{n}=\left\{\begin{array}{ll}
C \cdot \tanh\left(\frac{\mathbf{q}^{T}\mathbf{k}_{n}}{\sqrt{d_\text{h}}}\right) &,\text{ if } n\in \bar{\mathcal{N}}, \\
-\infty & ,\text { otherwise},
\end{array}\right. 
\label{compa}
\end{equation}
where $\mathbf{q}=\mathbf{W}_{\text{Q}} \hat{\mathbf{c}}$ and $\mathbf{k}_{n}=\mathbf{W}_{\text{K}} \mathbf{r}_{n}$ with trainable matrices $\mathbf{W}_{\text{Q}}\in \mathbb{R}^{d_\text{h}\times d_\text{h}}$ and $\mathbf{W}_{\text{K}}\in \mathbb{R}^{d_\text{h}\times d_\text{h}}$, and the compatibilities are clipped within $\left[-C,C\right]$ with a hyper-parameter $C$. Again, we use minus infinity to mask the SPs that have been already selected or violate the distance constraints. Finally, the conditional probability is obtained by normalizing the compatibilities as
\begin{equation}
p(a_{m(t)}=n \mid \mathcal{A}_{t-1}, \mathbf{h})=\frac{e^{u_{n}}}{\sum_{n'=1}^{N} e^{u_{n'}}},~~\forall n \in \mathcal{N}.
\label{10}
\end{equation}
Correspondingly, $a_{m(t)}$ is sampled from (\ref{10}) during the training process to enhance exploration.

\subsection{Design of $\mathcal{F}_\text{w}(\cdot,\cdot)$}
We first exploit the optimal solution structure~\cite{OptimMUMISO}:
\begin{equation}
\label{MAbeamstruc}
\begin{aligned}
{\mathbf{w}}_{k}=&\sqrt{p_{k}} \frac{\left(\mathbf{I}_{M}+\sum\limits_{i=1}^{K} \frac{\mu_{i}}{\sigma_{k}^{2}} \mathbf{h}_{i}(\mathcal{A}) \mathbf{h}_{i}(\mathcal{A}) ^{H}\right)^{-1} \mathbf{h}_{k}(\mathcal{A})}{\left\|\left(\mathbf{I}_{M}+\sum\limits_{i=1}^{K} \frac{\mu_{i}}{\sigma_{k}^{2}} \mathbf{h}_{i}(\mathcal{A}) \mathbf{h}_{i}(\mathcal{A}) ^{H}\right)^{-1} \mathbf{h}_{k}(\mathcal{A})\right\|}, \\
&~\forall k \in \mathcal{K},
\end{aligned}
\end{equation}
where $\boldsymbol{\mu}\triangleq[\mu_1,\mu_2,\dots,\mu_{K}]^T\in\mathbb{R}_{+}^{K}$ and $\boldsymbol{p}\triangleq[p_1,p_2,\dots,p_{K}]^T\in\mathbb{R}_{+}^{K}$ are unknown parameters and should satisfy $\sum_{k=1}^{K}\mu_{k}=\sum_{k=1}^{K}p_{k}=P_{\text{max}}$ such that \eqref{power} can be satisfied. Exploiting this solution structure can simplify the mapping to be learned by the NN, by decreasing the number of unknown parameters from $2KM$ to $2K$. 

The remaining task is to predict these low-dimensional parameters. For this purpose, we adopt the ENGNN, which consists of two types of nodes, i.e., $M$ MA-nodes and $K$ UE-nodes, and there exists an edge between each MA-node and each UE-node. The edge features are initialized as
\begin{equation}
\begin{aligned}
\hat{\mathbf{e}}^{[0]}_{mk}& =\text{MLP}^{\text{W}}_{1}(\Re\{[\mathbf{h}_{k}(\mathcal{A})]_m\},\Im\{[\mathbf{h}_{k}(\mathcal{A})]_m\}), \\
& ~\forall k \in\mathcal{K},m\in \mathcal{M},
\end{aligned}
\end{equation}
and the node features $\{\hat{\mathbf{f}}^{[0]}_{\text{MA},m}\}_{m\in\mathcal{M}}$ and $\{\hat{\mathbf{f}}^{[0]}_{\text{UE},k}\}_{k\in\mathcal{K}}$ are initialized as zero. Similar to \eqref{ENGNNUE}-\eqref{ENGNNEdge}, after $L_{\text{W}}$ layers' update, we obtain $\boldsymbol{\mu}$ and $\boldsymbol{p}$ by
\begin{subequations}
\begin{align}
&[{\mu}_k,{p}_k]^T= \text{FC}(\hat{\mathbf{f}}^{[L_{\text{W}}]}_{\text{UE},k}),~~\forall k \in \mathcal{K}, \\
&\boldsymbol{\mu} \gets P_{\max}\cdot\text{softmax}({\boldsymbol{\mu}}), \\
&\boldsymbol{p} \gets P_{\max}\cdot\text{softmax}({\boldsymbol{p}}), 
\end{align}
\end{subequations}
where $\text{FC}(\cdot)$ is a fully-connected layer, and the softmax activation is employed to normalize $\boldsymbol{\mu}$ and $\boldsymbol{p}$. Finally, $\boldsymbol{\mu}$ and $\boldsymbol{p}$ are substituted to (\ref{MAbeamstruc}) to get $\mathbf{w}$.

\subsection{Joint Training of $\mathcal{F}_{\text{p}}(\cdot)$ and $\mathcal{F}_{\text{w}}(\cdot,\cdot)$}
Based on the probabilistic modeling of $p(\mathcal{A}|\mathbf{h})$ in $\mathcal{F}_\text{p}(\cdot)$, the joint training problem can be formulated as
\begin{subequations}
\begin{align}
&\max_{\boldsymbol{\theta}_{\text{p}},\boldsymbol{\theta}_{\text{w}}}& &J(\boldsymbol{\theta}_{\text{p}},\boldsymbol{\theta}_{\text{w}})=\mathbb{E}_{\mathbf{h},\mathcal{A}\sim p(\mathcal{A}|\mathbf{h})}[R(\mathcal{A},\mathbf{w};\mathbf{h})], 
\tag{18}\label{DNNjointrain}
\end{align}
\end{subequations}
where $\boldsymbol{\theta}_{\text{p}}$ and $\boldsymbol{\theta}_{\text{w}}$ denote the trainable parameters of $\mathcal{F}_{\text{p}}(\cdot)$ and $\mathcal{F}_{\text{w}}(\cdot,\cdot)$, respectively, and the original constraints in \eqref{discrete_constraint}-\eqref{power} are satisfied through the design of $\mathcal{F}_\text{p}(\cdot)$ and $\mathcal{F}_\text{w}(\cdot,\cdot)$. Next, we show how to update $\boldsymbol{\theta}_\text{p}$ and $\boldsymbol{\theta}_\text{w}$.

\emph{1) Update of $\boldsymbol{\theta}_{\text{p}}$: }Since $\mathcal{A}$ is sampled according to $p(\mathcal{A}|\mathbf{h})$, $R(\mathcal{A},\mathbf{w};\mathbf{h})$ is non-differentiable with respect to $\boldsymbol{\theta}_\text{p}$. However, since $p(\mathcal{A}|\mathbf{h})$ itself is differentiable with respect to $\boldsymbol{\theta}_\text{p}$, we can calculate the gradient of $J(\mathbf{\boldsymbol{\theta}}_{\text{p}},\boldsymbol{\theta}_{\text{w}})$ with respect to $\boldsymbol{\theta}_{\text{p}}$ using the \emph{Policy-Gradient} strategy\cite{PolicyG}:
\begin{equation}
\begin{aligned}
&\nabla_{\boldsymbol{\theta}_{\text{p}}} J(\mathbf{\boldsymbol{\theta}}_{\text{p}},\boldsymbol{\theta}_{\text{w}}) = \mathbb{E}_{\mathbf{h}}\left[\sum_{\mathcal{A}\sim p(\mathcal{A}\mid \mathbf{h})} R(\mathcal{A},\mathbf{w};\mathbf{h}) \nabla _{\boldsymbol{\theta}_{\text{p}}}p(\mathcal{A} \mid \mathbf{h})\right] \\
&= \mathbb{E}_{\mathbf{h}}\left[\sum_{\mathcal{A}\sim p(\mathcal{A}\mid \mathbf{h})} R(\mathcal{A},\mathbf{w};\mathbf{h}) p(\mathcal{A}|\mathbf{h})\nabla _{\boldsymbol{\theta}_{\text{p}}}\log p(\mathcal{A} \mid \mathbf{h})\right] \\
&=\mathbb{E}_{\mathbf{h},\mathcal{A}\sim p(\mathcal{A}\mid \mathbf{h})}\left[R(\mathcal{A},\mathbf{w};\mathbf{h})
\nabla_{\boldsymbol{\theta}_{\text{p}}} \log p(\mathcal{A}\mid \mathbf{h})\right].
\end{aligned}
\label{pg}
\end{equation}
Then, $\boldsymbol{\theta}_\text{p}$ is updated to maximize (\ref{DNNjointrain}) by a mini-batch stochastic gradient ascent algorithm, which can be implemented with the Adam optimizer \cite{kingma2014adam}. 
\addtolength{\topmargin}{0.05in}
\begin{algorithm}[t]
\caption{Joint Training of $\mathcal{F}_{\text{p}}(\cdot)$ and $\mathcal{F}_{\text{w}}(\cdot,\cdot)$}
\begin{algorithmic}[1]
\STATE \textbf{Input} number of epochs $N^{\text{e}}$, steps per epoch $N^{\text{s}}$, and batch size $B$
\STATE \textbf{Initialize} $\boldsymbol{\theta}_{\text{p}}$ and $\boldsymbol{\theta}_{\text{w}}$
\FOR{epoch$=1,2,\dots,N^{\text{e}}$}
\FOR{step$=1,2,\dots,N^{\text{s}}$}
\STATE Select a batch of input $\{\mathbf{h}\}$ from the training dataset
    \STATE Output $p(\mathcal{A}|\mathbf{h})$, $\mathcal{A}$, and $\mathbf{w}$ corresponding to each training sample $\mathbf{h}$
    \STATE $\boldsymbol{\theta}_{\text{w}} \gets \text{Adam}(\boldsymbol{\theta}_{\text{w}},-\nabla_{\boldsymbol{\theta}_{\text{w}}}J(\boldsymbol{\theta}_{\text{p}},\boldsymbol{\theta}_{\text{w}}))$ 
\STATE $\boldsymbol{\theta}_{\text{p}} \gets \text{Adam}(\boldsymbol{\theta}_{\text{p}}, -\nabla_{\boldsymbol{\theta}_{\text{p}} }J(\mathbf{\boldsymbol{\theta}}_{\text{p}},\boldsymbol{\theta}_{\text{w}}))$
\ENDFOR
\ENDFOR
\end{algorithmic}
\label{algorithm1}
\end{algorithm}

\emph{2) Update of $\boldsymbol{\theta}_{\text{w}}$:} Since $R(\mathcal{A},\mathbf{w};\mathbf{h})$ is differentiable with respect to $\mathbf{w}$ and consequently with respect to $\boldsymbol{\theta}_\text{w}$, we can compute the gradient of $J(\mathbf{\boldsymbol{\theta}}_{\text{p}},\mathbf{\boldsymbol{\theta}}_{\text{w}})$ with respect to $\boldsymbol{\theta}_{\text{w}}$ by
\begin{equation}
\nabla_{\boldsymbol{\theta}_{\text{w}}} J(\mathbf{\boldsymbol{\theta}}_{\text{p}},\boldsymbol{\theta}_{\text{w}}) = \mathbb{E}_{\mathbf{h},\mathcal{A}\sim p(\mathcal{A}\mid \mathbf{h})}\left[\nabla_{\boldsymbol{\theta}_{\text{w}}}R\left(\mathcal{A},\mathbf{w};\mathbf{h}\right)\right].
\label{wtrain}
\end{equation}
Consequently, $\boldsymbol{\theta}_\text{w}$ is updated to maximize (\ref{DNNjointrain}) by a mini-batch stochastic gradient ascent algorithm using the Adam optimizer.

\begin{figure*}[htbp]
\centerline{\includegraphics[scale=0.32]{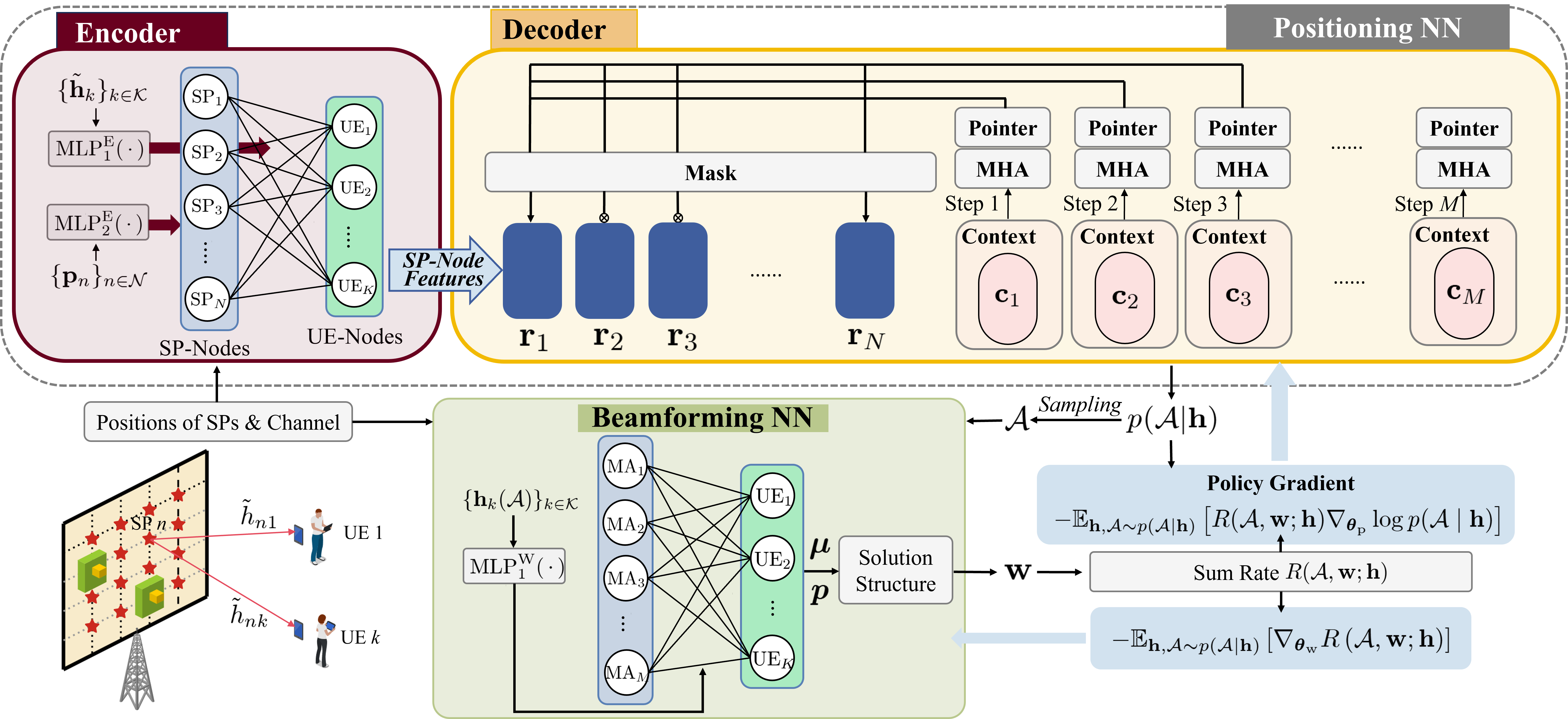}}
\caption{Structure of our proposed DL framework.}
\label{fig1}
\end{figure*}

The joint training algorithm is summarized in Algorithm~\ref{algorithm1}, where the negative gradient is adopted in lines 7 and 8 to set a gradient \emph{ascent} update in the Adam optimizer. After training, $\mathcal{F}_{\text{p}}(\cdot)$ and $\mathcal{F}_{\text{w}}(\cdot,\cdot)$ with parameters $\boldsymbol{\theta}_{\text{p}}$ and $\boldsymbol{\theta}_{\text{w}}$ are used to predict $\mathcal{A}$ and $\mathbf{w}$ given any $\mathbf{h}$. The proposed overall DL framework is summarized in Fig. \ref{fig1}.

\section{Numerical Results}
In this section, we evaluate the performance of the proposed DL framework via numerical results. We consider an MA-aided system where the BS is equipped with $[4,9]$ MAs to serve $4$ single-antenna UEs. The size of the 2D rectangular transmit area is $2\lambda\times2\lambda$, where each side is uniformly sampled with $[5,8]$ points, resulting in $\{25,36,49,64\}$ SPs, and the wavelength $\lambda=60\text{ mm}$. The minimum distance between every two MAs is set as $d_{\text{min}}= 30\text{ mm}$. The distance between the $k$-th UE and the BS, denoted as $D_k$, is uniformly distributed within $[100,200]$ in meters. The noise power is set as $-100$ dBm.

We consider the field-response channel model (see (3)-(6) in \cite{10318061}) where $\{\tilde{\mathbf{h}}_{k}\}_{k\in\mathcal{K}}$ is determined by the positions of SPs, the path-response matrix $\boldsymbol{\Sigma}_k\triangleq\operatorname{diag}\left\{\left[\eta_{1k},\eta_{2k}, \dots, \eta_{L_{\text{r}}k}\right]^{T}\right\}$, the elevation angle of departure (AoD) $\{\theta_{k,l_{\text{t}}}\}_{k=1,l_{\text{t}}=1}^{K,L_{\text{t}}}$, and the azimuth AoD $\{\phi_{k,l_{\text{t}}}\}_{k=1,l_{\text{t}}=1}^{K,L_{\text{t}}}$, where $L_{\text{t}}=L_{\text{r}}=16$ are the numbers of transmit paths and receive paths, respectively, and $\eta_{l_\text{r}k}$ follows complex Gaussian distribution $\mathcal{C N}\left(0, L_{0} D_{k}^{-\alpha}\right)$ with $L_0=34.5$ dB and $\alpha=3.67$. Besides, the probability density function of AoDs is $f_{\text{AoD}}\left({\theta}_{k, l_{\text{t}}}, \phi_{k, l_{\text{t}}}\right)={\cos {\theta}_{k, l_{\text{t}}}}/{2 \pi},~{\theta}_{k, l_{\text{t}}} \in\left[-{\pi}/{2},{\pi}/{2}\right],~\phi_{k, l_{\text{t}}} \in\left[-{\pi}/{2},{\pi}/{2}\right]$.\footnote{We assume that the elevation AoD $\{\theta_{k,l_{\text{t}}}\}_{k=1,l_{\text{t}}=1}^{K,L_{\text{t}}}$, azimuth AoD $\{\phi_{k,l_{\text{t}}}\}_{k=1,l_{\text{t}}=1}^{K,L_{\text{t}}}$, and the path response coefficients $\left\{\eta_{l_rk}\right\}_{k=1,l_{\text{r}}=1}^{K,L_{\text{r}}}$ are perfectly estimated \cite{MAChannelEstimation}, enabling the perfect recovery of the channel $\mathbf{h}$.}

For the proposed NN model, all the MLPs are implemented by 2 linear layers, each followed by a ReLU activation function. The encoder of $\mathcal{F}_{\text{p}}(\cdot)$ has $L_\text{E}=3$ layers, with $n_{\text{emb}}=128$ and $d_\text{h}=256$. The number of heads in the MHA model is $N_\text{h}=8$. The clipping logit is set to $C = 8$. Moreover, $\mathcal{F}_{\text{w}}(\cdot,\cdot)$ has $L_{\text{W}}=3$ hidden layers, each with a hidden dimension of 64. In the training procedure, the number of epochs is set to 100, where each epoch consists of 50 mini-batches with a batch size of $1024$. A learning rate $10^{-4}$ is adopted to update the trainable parameters $\boldsymbol{\theta}_{\text{p}}$ and $\boldsymbol{\theta}_{\text{w}}$ through Algorithm \ref{algorithm1} using the Adam optimizer. 
 
After training, we test the performance of the proposed DL framework on a testing set comprising 1000 samples. The implementation of the experiment is under the PyTorch version 2.1.2+cu12, operating on an NVIDIA GeForce GTX 4090 GPU. 

We consider four baselines for comparison:
\begin{itemize}
\item \textbf{Random+WMMSE:} This method involves randomly selecting $M$ elements from $\mathcal{N}$ to construct $\mathcal{A}$. If $\mathcal{A}$ violates the constraints in (\ref{distance_constraint}), the selection is repeated until the constraints are satisfied. Subsequently, with the MA positions determined, $\mathbf{w}$ is computed using the WMMSE algorithm\cite{WMMSE}.
\item \textbf{Strongest+WMMSE:} For this approach, the average channel gain of each SP with respect to all $K$ UEs is first calculated and treated as the equivalent channel gain for that SP. The SPs exhibiting the strongest channel gains are then selected iteratively over $M$ steps, with SPs violating (\ref{distance_constraint}) masked in each step. Following the positioning, $\mathbf{w}$ is obtained via the WMMSE algorithm.
\item \textbf{Strongest+ZF:} This method combines strongest-channel-based positioning with the zero-forcing beamforming \cite{ZF}.
\item \textbf{FP-C:} We modify an approach that jointly optimizes continuous antenna positioning and beamforming for MA-aided systems \cite{C_FP_MA}. To obtain discrete antenna positions that satisfy the discrete constraints in \eqref{distance_constraint}, in each iteration of FP-C we project the position of each MA to its nearest discrete SP that satisfies \eqref{distance_constraint} sequentially.
\end{itemize}
\begin{figure}[htbp]
\centering
\includegraphics[scale=0.05]{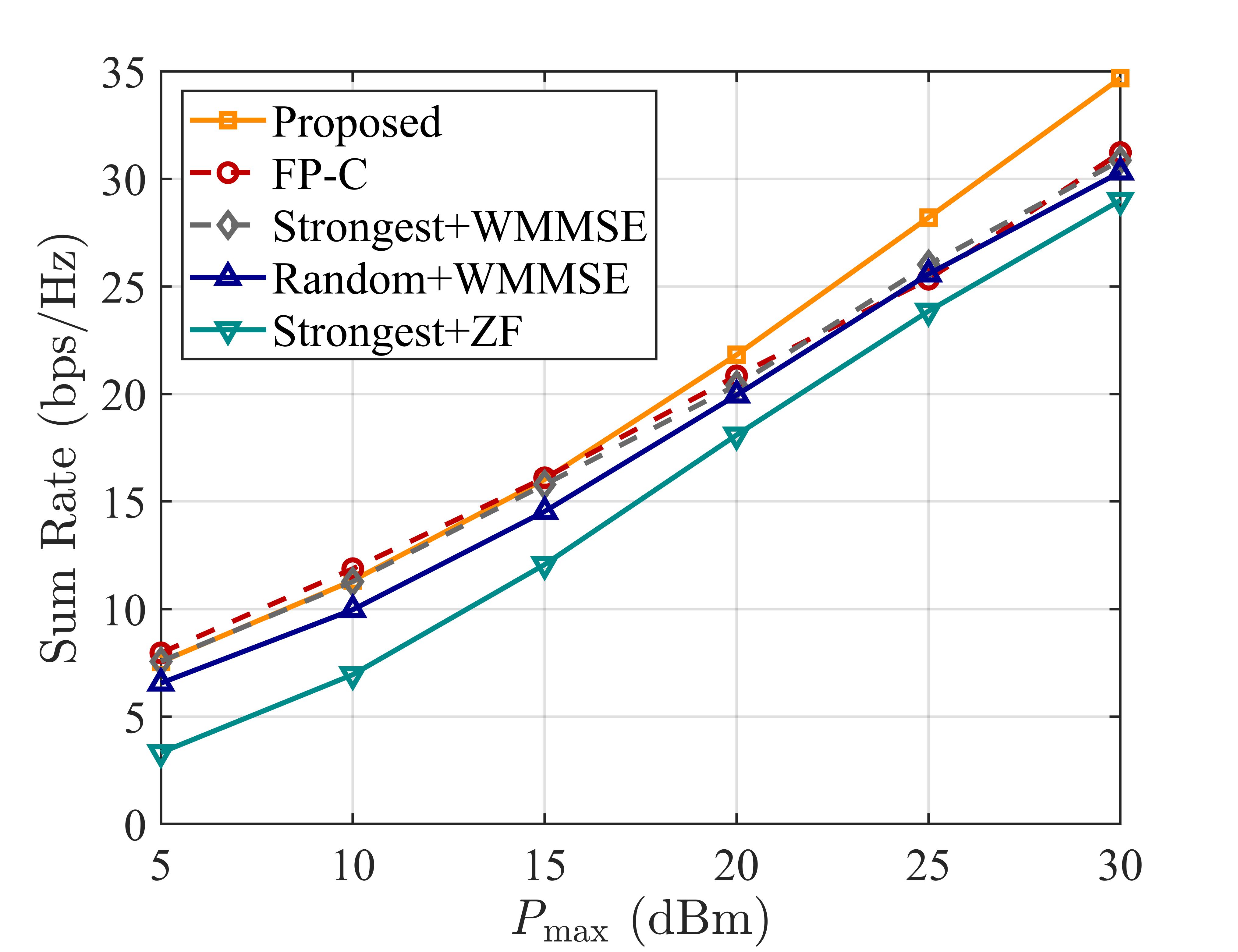}
\caption{The sum rate performance comparison under different power budgets when $M=6$ and $N=49$.}
\label{MAFig2}
\end{figure}

We first compare the sum rate performance under different power budgets. In Fig.~\ref{MAFig2}, with $M=6$ and $N=49$, it can be observed that \emph{Random+WMMSE} and \emph{Strongest+ZF} perform badly due to their heuristic antenna positioning and beamforming strategies, respectively. Moreover, the proposed DL framework outperforms all baselines particularly at higher transmit power. This demonstrates the great capability of the proposed framework in mitigating interference, particularly when inter-user interference becomes stronger. 
\begin{table}[h]
\caption{Comparison of Average Computation Time}
\centering
\begin{tabular}{lr}
\toprule
Method & Time (ms) \\
\midrule
Strongest+ZF & 3.26  \\
Random+WMMSE & 644.37 \\
Strongest+WMMSE & 644.66 \\
FP-C & 3682.82 \\
Proposed & 7.89 \\
\bottomrule
\end{tabular}
\label{MAtable1}
\end{table}

Next, we compare the average computation time across various approaches. It can be observed from Table~\ref{MAtable1} that the computation time of the proposed DL framework is substantially shorter than the iterative optimization-based methods, including \emph{WMMSE} and \emph{FP-C}.
\begin{figure}[H]
\centering
\subfigure[Under different $M$'s.]{\includegraphics[width=0.24\textwidth]{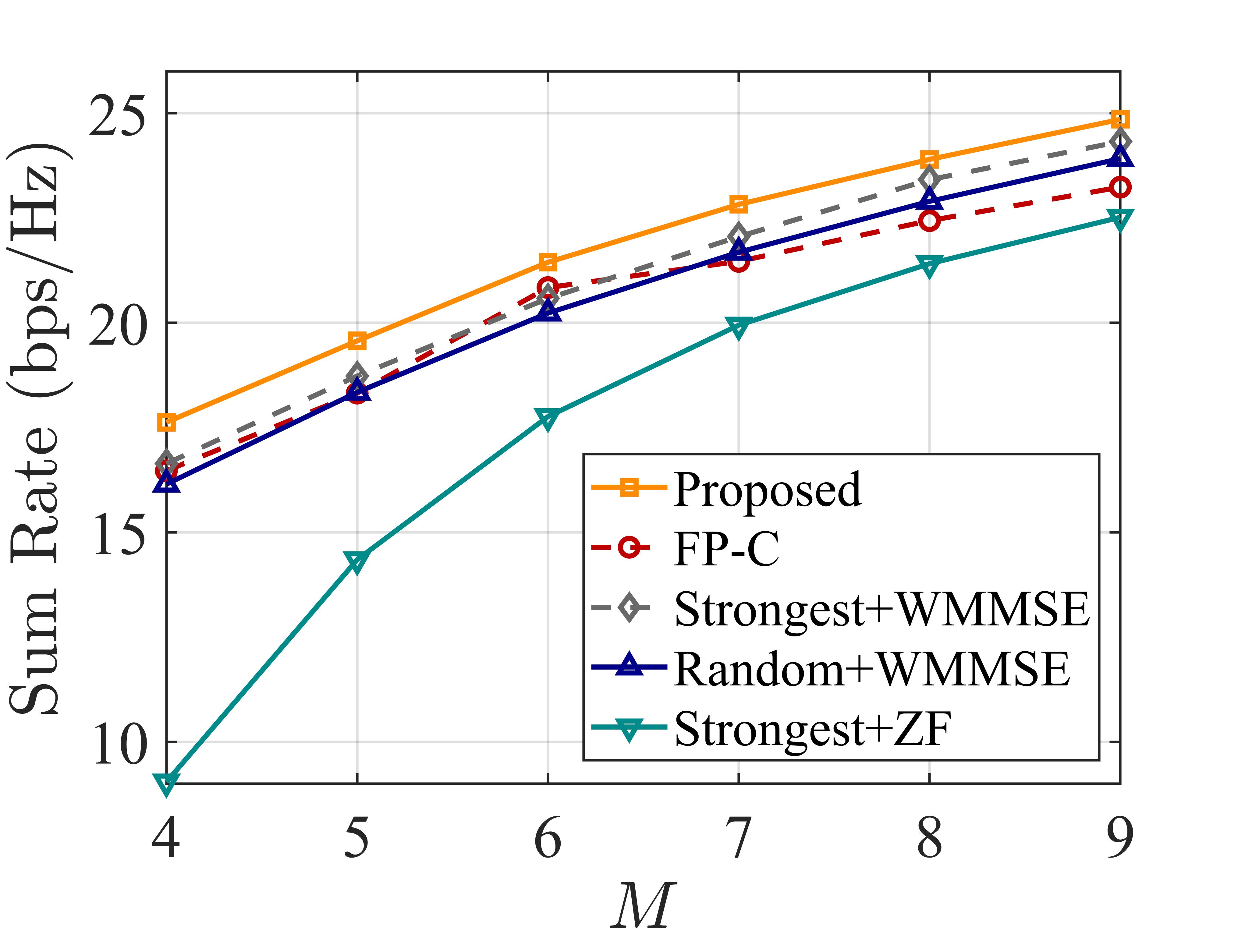}%
\label{4sum_rate}}
\subfigure[Under different $N$'s.]{\includegraphics[width=0.24\textwidth]{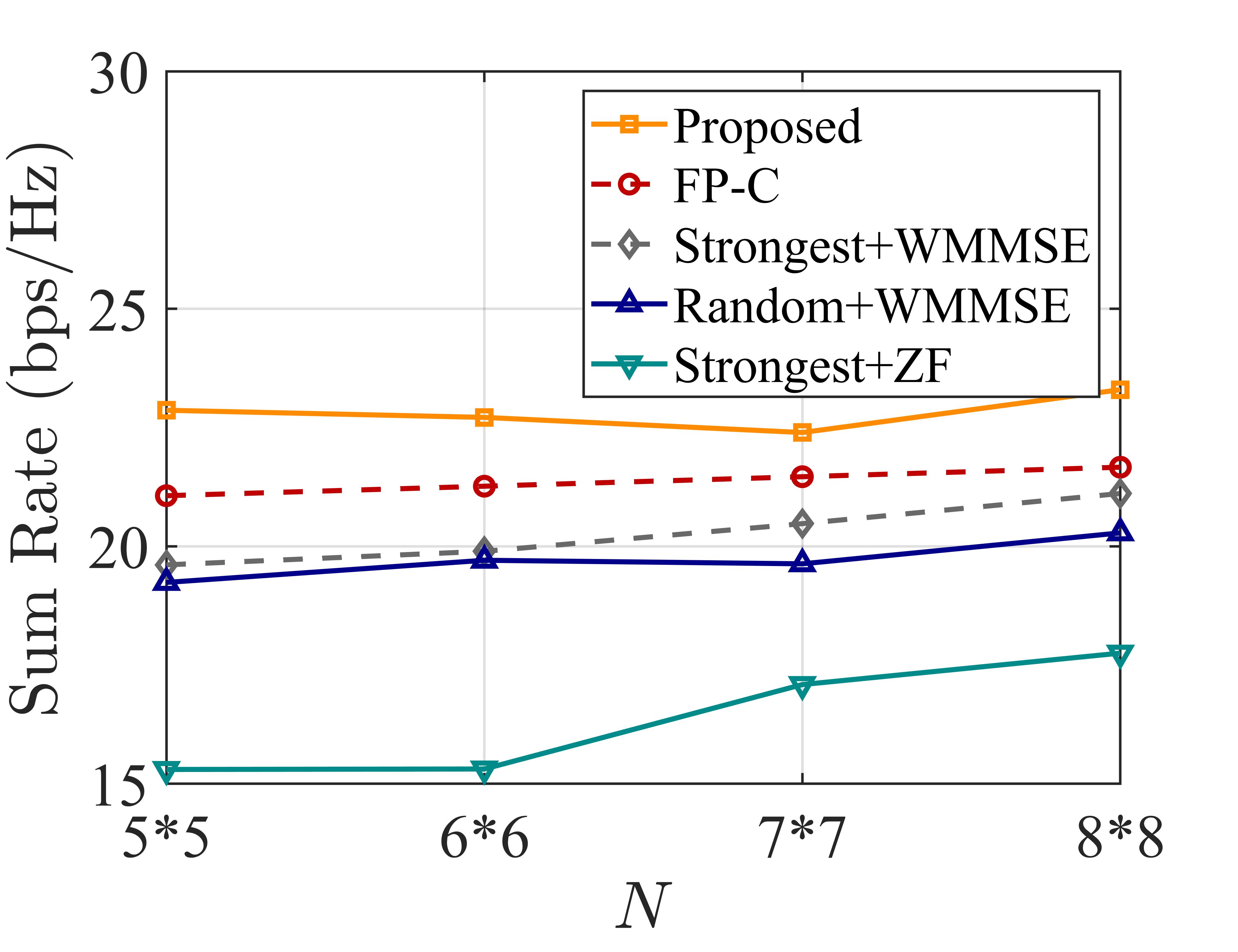}%
\label{5sum_rate}}
\caption{The sum rate performance of our proposed DL framework under different $M$'s and $N$'s.}
\label{MAFig3}
\end{figure}

Finally, we evaluate the performance of the proposed DL framework across a range of system settings. We first set $P_{\text{max}}=20$ dBm and $N=49$, and plot the sum rate against $M$ in Fig.~\ref{MAFig3}(a).  As observed, the proposed DL framework achieves highest sum rates across different values of $M$. Next, in Fig.~\ref{MAFig3}(b), we plot the sum rate against $N$ when $P_{\text{max}}=20$ dBm and $M=6$. As observed from Fig.~\ref{MAFig3}(b), the proposed DL framework achieves superior performance across different values of $N$. In contrast to heuristic positioning methods such as \emph{Strongest} and \emph{Random} which perform poorly, our proposed method achieves the best performance no matter when the number of SPs is limited, or when the number of SPs is sufficient but with more challenging constraints.

\section{conclusion}
In this paper, we have proposed a novel DL framework for solving the discrete antenna positioning and beamforming design problem in MA-aided multi-user systems. First, an encoder-decoder-based positioning NN has been developed to determine the MA positions, incorporating a sequential decoding strategy with a mask design to handle the discrete variables and the coupled distance constraints. Subsequently, the continuous variables are optimized using a beamforming NN, which leverages an ENGNN model informed by an optimal solution structure. Furthermore, we have introduced a joint training algorithm to jointly optimize the positioning NN and the beamforming NN. Numerical results have shown that the proposed end-to-end DL framework outperforms baseline approaches with much faster computation speed.
\bibliographystyle{IEEEtran}
\bibliography{ref}
\end{document}